\documentclass[twocolumn,prl]{revtex4}
\usepackage{graphicx}
\usepackage[dvips]{hyperref}
\usepackage{dcolumn}

\begin{document}
\def\simlt{\lower.5ex\hbox{$\; \buildrel < \over \sim \;$}}
\def\simgt{\lower.5ex\hbox{$\; \buildrel > \over \sim \;$}}
\def\simpt{\lower.5ex\hbox{$\; \buildrel \propto \over \sim \;$}}
\def\mag{\mbox{ mag}}
\def\kms{\mbox{ km s$^{-1}$}}
\def\mpc{\mbox{ Mpc}}
\def\kpc{\mbox{ kpc}}
\def\pc{\mbox{ pc}}
\def\deg{\mbox{ deg}}
\def\yr{\mbox{ yr}}
\def\msun{\mbox{ M}_\odot}
\def\clight{\mbox{c}}

\def\mnras{MNRAS}
\def\apj{ApJ}
\def\apjl{ApJL}
\def\aj{AJ}
\def\aap{AAP}

\title{New Constraints on Macroscopic Compact Objects as a Dark Matter
  Candidate from Gravitational Lensing of Type Ia Supernovae}

\date{\today}

\author{R. Benton Metcalf}
\affiliation{\it Max Plank Institut f\"ur Astrophysics, Karl-Schwarzchild-Str. 1, 85741 Garching, Germany}

\author{Joseph Silk}
\affiliation{\it Astrophysics, University of Oxford, Denys Wilkinson Building, Keble Road, Oxford OX1 3RH, UK}

\begin{abstract}
  We use the distribution, and particularly the skewness, of high
  redshift type Ia supernovae brightnesses relative to the low
  redshift sample to constrain the density of macroscopic compact
  objects (MCOs) in the universe.  The data favors dark matter made of
  microscopic particles (such as the LSP) at 89\% confidence. Future data will greatly improve this limit.  This constraint is valid for a range in MCO mass from $10^{-2}\msun$ to $10^{10}\msun$.  Combined with other constraints, MCOs larger than one tenth the mass of Earth ($\sim 10^{-7}\msun$) can be eliminated as the sole constituent of dark matter.   
\end{abstract}


\maketitle
  
The recent observational success of the cold dark matter (CDM) model puts only loose constraints on the properties of  dark matter itself.  
The requirements for a cold dark matter candidate can be satisfied by a microscopic, weakly interacting particle or by macroscopic objects that do not produce significant amounts of observable radiation through astrophysical processes, such as primordial black holes (PBHs).  
Direct observational constraints are needed.

There have been a number of previous constraints put on the density of massive compact objects (MCOs) from a variety of observations.  
The rate of gravitational microlensing of stars in the Small and Large Magellanic
Clouds puts a limit on the mass fraction of MCOs in our galaxy's dark matter
halo of $< 40\%$ with masses between $10^{-7}$ and $10\msun$  assuming a
"standard" halo model
\citep{astro-ph/0607207,2000ApJ...542..281A,1998ApJ...499L...9A}.
Microlensing toward M31 has thus far been inconclusive \citep{2006A&A...446..855D,2005A&A...443..911C}. 
From the effect of gravitational lensing on the equivalent widths of
quasar broad lines \cite{1994ApJ...424..550D}  placed a limit of
$\Omega_{\rm mco}< 1$ for $10^{-3} \msun \simlt M \simlt 300\msun$. 
Using Very Large Baseline Interferometry (VLBI), \cite{2001PhRvL..86..584W} looked for radio QSOs that are gravitationally lensed into multiple images.  They put an upper limit of $\Omega_{\rm mco}<0.01$ for very massive ($10^6 - 10^{8}\msun$) MCOs.
There are also arguments from stellar dynamics for why such high mass MCOs cannot make up a significant fraction of the Milky Way's dark matter halo
\citep{1993ApJ...413L..93M,1985ApJ...299..633L}. 
 Although these arguments eliminate the possibility of all the dark matter being in this form it is difficult to be more quantitative since we do not have detailed information on the initial population of globular clusters or the velocity dispersion of the disk stars.  All of  these constraints leave a large open window between $\sim 100\msun$ and $10^5\msun$ which this study seeks to eliminate.

Type Ia supernovae are found empirically to be good standard candles with luminosities large enough to be observable at cosmological distances.  They have been used to measure the expansion rate of the universe as a function of time and through this the densities of matter and dark energy \citep[][and references therein]{2003ApJ...598..102K,2004ApJ...607..665R}.  
The distribution of supernova brightnesses will also be systematically affected by the gravitational fields along the line of sight.   In the most extreme case where all the matter in the universe is in macroscopic objects, the magnification distribution will be highly skewed with most high redshift supernovae appearing less bright than they would in a perfectly homogeneous universe and a small number being highly magnified \citep{1991ApJ...374...83R,HW98,1999ApJ...519L...1M,1999A&A...351L..10S}.  In this paper we will use this effect to measure what fraction of dark matter can be made of MCOs.  
The systematic skew in the distribution of luminosity distances can also affect the implied best-fit cosmological parameters. The cosmological constant is likely to be overestimated if MCOs exist in large numbers but are not included in the data analysis.  We show how significant this mis-measurement could be.

Regardless of what the basic unit of dark matter is,  SNe will be weakly lensed by the clumping of the dark matter and baryons into galaxies, dark matter halos and larger scale structures \citep{1999MNRAS.305..746M,2001MNRAS.327..115M}.  However this effect is relatively weak and as yet undetected \cite{2005MNRAS.358..101M} (see \cite{2004MNRAS.351.1387W} for a contrary opinion).
We choose the conservative option of not including this type of weak lensing in our analysis.  We derive an upper bound on the density of MCOs and including clumping would only reduce this bound at the expense of further assumptions.

With the use of the Sachs optical scalar equations \citep{sachs61} the
cross section of a beam of light can be calculated as a function of
the affine parameter along a fiducial light path, $\lambda$.  The beam is affected by both the local
mass-energy density that the beam passes through, represented by the
Ricci tensor, and by nonlocal, inhomogeneous shear contributions coming from mass
outside of the beam which are proportional the the Weyl tensor.
Ignoring the Weyl contribution, 
the angular size distance can be define as the solution to the
following equation and boundary conditions \citep{1973ApJ...180L..31D} 
\begin{eqnarray}\label{eq:distance}
\frac{1}{D}\frac{d^2 D}{d\lambda^2}=  -\frac{4\pi G}{c^2} \rho(\lambda) (1+z)^2 \\
D|_{\lambda=0}=0~~,~~\left.\frac{d D}{d\lambda}\right|_{\lambda=0} = 1.  \nonumber
\end{eqnarray}
It has been assumed that the light beam passes through a
fluid of density $\rho(\lambda)$ and negligible pressure ($p<<\rho c^2$).
The affine parameter is related to the redshift by $d\lambda=
c H_o^{-1}(1+z)^{-2}{\rm E}(z)^{-1} dz$ where ${\rm E}(z)=\sqrt{\Omega_m (1+z)^3 + \Omega_\Lambda +
  (1-\Omega_m-\Omega_\Lambda)(1+z)^2}$.

The standard angular size distance calculated with the
Friedman-Robertson-Walker (FRW) metric (the angular size distance in a
perfectly homogeneous universe), here denoted by
$\overline{D}(z,\Omega_m,\Omega_\Lambda)$, is a solution to 
(\ref{eq:distance}) in the special case of $\rho(z)=\overline{\rho}(z)$
where $\overline{\rho}(z)$ is the average density of the universe.
The angular size distance appropriate 
for a light beam that travels through empty space or through a density
of matter that is less than the average density will be larger than
the FRW distance.  The angular size distance between
two redshifts $z_1$ and $z_2$ for a beam that passes
through a fixed fraction of the average density, $f_{\rm mco}$,  will be
denoted $D(z_1,z_2,f_{\rm  mco},\Omega_m,\Omega_\Lambda)$.  We call this the {\it unfilled beam distance} when $f_{\rm mco} < 1$.

If a point-like source is being lensed by a single point-like mass, the source
position on the lens plane, $y$, can be expressed in terms of the
magnification $y=\sqrt{2}R_e(z_s,f_{\rm mco},\Omega_m,\Omega_\Lambda)\sqrt{\mu(\mu^2-1)^{-1/2}-1}$ \citep{1991ApJ...366..412G}.  There are two images of the source and $\mu$ is the sum of their magnifications.  The Einstein radius is given by 
\begin{eqnarray}
R_e(z,z_s,f_{\rm mco},\Omega_m,\Omega_\Lambda) = \sqrt{\left(\frac{4GM}{c^2}\right) \frac{D(0,z)D(z,z_s)}{D(0,z_s)}}.
\end{eqnarray}
The distances that appear in the Einstein radius are the unfilled beam
distances and not $\overline{D}(z)$. 
The magnification probability distribution function (pdf)  is easily calculated
\begin{eqnarray}\label{1_lens_prob}
P(\mu)~d\mu & = & 2\pi
\int_0^{z_s}dz~\frac{d\chi}{dz}\frac{\rho_{\rm mco}(z)}{M} y
\frac{dy}{d\mu} d\mu \\ \nonumber
& = & 2\, \,
\frac{\tau\left(z_s,f_{\rm mco},\Omega_m,\Omega_\Lambda\right)}{\left| \mu^2 -1 \right|^{3/2}} ~d\mu 
\end{eqnarray}
where $\chi$ is the coordinate distance in the radial direction.
The normalization, $\tau\left(z_s,f_{\rm mco},\Omega_m,\Omega_\Lambda\right)$, is often called the optical depth.  
Note that the optical depth depends only on the average mass density
of MCOs and not on the mass of the individual MCOs.  It is also
independent of the Hubble parameter since $D\propto H_o^{-1}$.

The magnification pdf for one lens is not sufficient for an
ensemble of point-like lenses.  
There is no way of analytically calculating the full pdf for an ensemble of lenses.
However, there are several requirements that the full pdf must fulfill.
One is that the average angular size distance to objects at a fixed
redshift be the same as it would be in a homogeneous universe
\citep{2005ApJ...632..718K}.  This is the requirement that
$\langle\mu\rangle D(z)^{-2} = \overline{D}(z)^{-2}$.  
We also expect that for $\mu<1$ the pdf is zero or at least very small.  This is because when the optical depth is small (as it is for all the cases considered here) the $\mu \simeq 1$ cases occur for lines of sight that pass well away from any lens where there will be only one significant image.  The weak lensing limit ($ \int dr D(0,z)D(z,z_s)\, \vec{\bigtriangledown}\vec{\bigtriangledown} \phi(r) \ll c^2D(0,z_s)$, where $\phi$ is the Newtonian potential and the integral is in the radial direction) will apply in these cases and in this limit there must be at least one image with $\mu\ge 1$ \citep{1984A&A...140..119S}. 

In addition, we expect that when the optical depth is small  the
true pdf will have the same form at large $\mu$ as (\ref{1_lens_prob})
because in these rare cases the line of sight passes very close to one
lens and is unlikely to pass very close to two or more. 
Numerical simulations give a magnification pdf with a high $\mu$ tail very similar to
(\ref{1_lens_prob}), as expected, and a strong, but not singular mode
at $\mu=1$ \citep{1991ApJ...374...83R,HW98,1999ApJ...519L...1M}.  In this paper, 
to  estimate the magnification pdf we use an analytic form
advocated by \cite{1991ApJ...374...83R}:
\begin{eqnarray}\label{full_pdf}
P(\mu;\delta,\tau_{\rm eff})d\mu = \left\{ 
\begin{array}{ccl}
2\, \tau_{\rm eff} \left( \frac{1 -
  \exp\left(\frac{1-\mu}{\delta}\right) }{\mu^2 -1} \right)^{3/2} d\mu &
, & \mu \ge 1 \\
0 & , & \mu < 1
\end{array}
\right.
\end{eqnarray}
We fix the two free
parameters, $\delta$ and $\tau_{\rm eff}$, 
by applying the average magnification constraint and the normalization
constraint numerically.
We have experimented with other forms for the pdf and find that any
pdf that satisfies the expectations given above and is monotonic and
reasonably smooth will give very similar results for the density of
MCOs. 

We use two sets of publicly available SN data.
Firstly, the Riess {\it et al.} \cite{2004ApJ...607..665R} (here after RST) ``gold'' sample containing 157 type Ia SNe.
 There is an empirical relationship between the width of
the SN's light curve and the peak luminosity of the SN.  They derive template light curves in different passbands
from an independent sample of SNe \citep{2006AJ....131..527J} and then fit the light curves of the high redshift SNe to these templates to determine a corrected peak luminosity for each SN. 
Because the correction method is derived from a low redshift sample of SNe that should not be affected by gravitational lensing it can be justifiably used to look for lensing in the high redshift sample.

Secondly, the Supernova Legacy Survey (SNLS) \citep{AstierData} 
uses 44 "nearby" SNe ($0.015<z\leq0.125$) from the literature, some of which are included  in the RST gold sample, plus 73 of their own high redshift SNe ($0.249\leq z \leq 1.01$).
SNLS makes the peak-luminosity correction by the simple relation
$m_{ob} = m^*_{B} +\alpha (s-1) -\beta c$
where $s$ is the "stretch", a measure of the light curve width, $c$ is the color ($=B-V+0.057$) and $m^*_{B}$ is the observed peak magnitude in rest frame $B$-band.  These three parameters are fit to each observed light curve.  The parameters $\alpha$, $\beta$ and the absolute peak magnitude of the SNe are fit to the whole data set along with the cosmological model.  In what follows we refit these parameters since the fits done by SNLS were done without including the possibility of gravitational lensing.

When gravitational lensing is included, the probability distribution for the magnitude of a SN is not Gaussian and $\chi^2$ is not a useful statistic.  We instead adopt a Bayesian approach to constrain the parameters.
For each SN we use a likelihood function of the form
\begin{eqnarray}
\begin{array}{rl}
{\mathcal L} &=  \int_{-\infty}^{\infty} d\mu~
 P_\mu\left(\mu|z_s\right)  P_m( m_{ob},\sigma_{ob},\mu)
\end{array}
\end{eqnarray}
where $m_{ob}$ is the observed, light-curve-corrected magnitude, $\sigma_{ob}$ is the reported observational error.

It is probable that the intrinsic distribution of corrected peak luminosities  is not Gaussian distributed.  If the intrinsic distribution has a skew it could be
mistaken for a lensing effect (if negative) or it could obscure a lensing signal (if positive).   To account for this we allow the intrinsic distribution of absolute magnitudes to have a redshift independent skewness.  This is implemented through modeling $P_m$ with a truncated Edgeworth expansion 
\begin{eqnarray}
P_m=\frac{e^{-x^2/2}}{\sqrt{2\pi(\sigma_m^2+\sigma_{ob}^2)}}  \left( 1+\frac{\kappa_3}{6} x(x^2-3) \right) \\ \nonumber
x=\frac{\left(m_{ob} - 5\log(D_L) + 2.5  \log(\mu)-M_o \right)}{\sqrt{\sigma_m^2+\sigma_{ob}^2}}
\end{eqnarray}
where $\kappa_3$ is the skewness.  The lower redshift SNe, which are not lensed, constrain $\kappa_3$.  This gives a total of 6 free parameters for the RST SNe. For SNLS SNe there are two additional parameters ($\alpha$ and $\beta$).

Recent CMB observations and galaxy surveys strongly indicate that the universe is flat ($\Omega_m+\Omega_\Lambda=1$) or very close to it \citep{2004PhRvD..69j3501T}.  In this paper we will restrict
ourselves to the flat case.    We also
use an independent constraint on $\Omega_m$ from WMAP and the HST Key project measurement 
of the Hubble parameter as reported in \cite{wmap3year}.  Specifically $\Omega_m = 0.28 \pm 0.04$.  This constraint is implemented as a Gaussian prior in the likelihood function.  For all other parameters we use uniform priors.

To find the confidence regions for all the parameters we perform a Monte Carlo Markov Chain  (MCMC) calculation.  This allows use to calculate a constraint on $f_{\rm mco}$ (or any other parameter) that is marginalizing over all the other parameters.   
In each MCMC at least 60,000 combinations of the parameters are tested.  
Two sets of calculations are done -- one in which $f_{\rm sub}$ is allowed to take any value between 0 and 1 and a second set in which $f_{\rm sub}$ is either 0 or 1.

The recovered values for $\alpha$, $\beta$, the $\sigma_m$'s and the $M_o$'s are all in good agreement with what was reported in the data papers.  Considering first the comparison between the two hypotheses $f_{\rm sub}=0$ and 1, we find that the SNLS data proffers $f_{\rm sub}=0$ at 89\% confidence.   The RST data proffers $f_{\rm sub}=1$ at 84\% confidence.  These seem in contradiction until the $\kappa_3$ distributions are considered.  For the RST data there is a strong positive correlation between $\kappa_3$ and $f_{\rm sub}$ while for the SNLS data they are not correlated.   If $\kappa_3$ is fixed at 0 the RST data shows no preference between $f_{\rm sub}=0$ and 1 while the preference persists in the SNLS data.   A maximum likelihood comparison test shows that including $\kappa_3 \neq 0$ does not improve the fit significantly enough that it is required by the data.  For this reason we do not think that the RST data contains enough information on the skewness of the brightness distribution to be useful for constraining $f_{\rm sub}$.
 This difference is probably a result of the different methods for correcting the peak luminosity.

\begin{figure}[t] 
   \includegraphics[width=10cm]{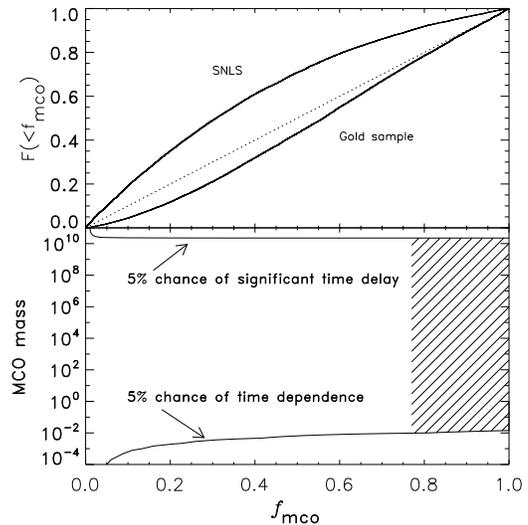} 
\caption[the]{\footnotesize 
On top is the Bayesian probability that the MCO fraction is less than $f_{\rm
    mco}$.  The top curve is for the SNLS SN sample and the bottom is for the gold sample of \cite{2004ApJ...607..665R}.  
    The dotted line is $F(<f_{\rm mco})=f_{\rm mco}$.
    In the bottom panel
    the two solid curves show where the approximation of point-like source and zero time delay begin to breakdown and time-dependent lensing effects are expected.  
    The hashed region is ruled out at 90\% c.l. by the SN data plus the WMAP constraint on $\Omega_m$.  
}
\label{fig:likeMCMC}
\end{figure}

Constraints on continuous values for $f_{\rm mco}$ are shown in the
top panel of figure~\ref{fig:likeMCMC}.  The SNLS puts a Bayesian
upper limit on $f_{\rm mco}$ of 88\% at a 95\% confidence level. 
 
To see if including the possibility of MCOs in the analysis changes the best fit $\Omega_m$ we repeat the calculations without the prior constraint on $\Omega_m$.  The best fit $\Omega_m$ for the SNLS data does not change significantly.
There is a larger although still small effect for the RST data ($\Omega_m = 0.29\pm 0.03$ to $0.31\pm 0.05$ when the prior is removed).

So far the SNe have been treated as  perfect point sources.  
Since
a SN's photosphere does have a finite size there
is a lower mass bound where this approximation fails to hold.  This
comes about when the source size, projected onto the lens plane $l_s$,
is of order $R_e$ or smaller and the impact parameter $y$ is of order
the sources size or smaller \citep{SW87,LSW88}.
We calculate the magnification numerically approximating the surface brightness profile as a
uniform disk (the photosphere is expected to be optically thick during these stages of the explosion)  and take a
deviation of 0.5~mag from the point source value to be significant.
This requirement puts a upper limit on $y$ as a function of $l_s/R_e$
for a lens that could violate the point source approximation.  Using
this we calculate the probability that there is a lens that meets
these conditions in the same manner that $\tau(z_s,f_{\rm mco})$ was
calculated previously.

The expansion rate of type Ia SN photospheres is
$v_{\rm phot} = 1.0-1.4\times10^4 \kms$ \citep{Patat96}.  We take the 
 typical period in which a SN is observed to be 20~days.  If the
 SN reaches a size where the point source approximation is violated
 during this period the SN's light curve will be significantly affected.  In
 the bottom panel of figure~\ref{fig:likeMCMC} the curve is shown where there is a 5\% chance that more than 1\%
 of the SNe are affected in this way.  The SN redshift distribution is taken to be the same as in the RST sample.
  This can be considered a conservative lower mass limit
 on where our constraints on $f_{\rm mco}$ are valid.

The magnification can also become time-dependent when the MCO mass scale
is very large because of the time delay between the gravitationally lensed images.  If this time delay is large enough compared to
the width of the peak in the light curve and the images have similar 
magnifications then the light curve will be affected.  
The rise time for a type Ia SN is much shorter than the fall time so the fractional change in the brightness,$\delta f$, reaches a maximum when the trailing image reaches the maximum.  For the purposes of calculation we approximate the light curve as exponentially falling from its maximum with a time scale of 20~days.  We consider a significant change in the light curve to be $\delta f =0.06$ corresponding to 3 magnitudes below the peak magnitude.

For a lens to cause a change in the light curve larger than $\delta f$ the SN must be within a range in $x$ where the outer limit is set by when the magnification of the trailing image is too small to cause a significant change and  the inner radius is set by the requirement that the time delay be large enough relative to the fall time of the SN.  For a small enough mass it is not possible to satisfy both of these requirements at the same time.
The probability of a lens being in this range can be calculated in a similar way to the calculation of $\tau(z_s,f_{\rm mco})$.  The curve where there would be a 5\% chance of having a significant change in the light curve of at least one of the SNe in the RST sample is shown in the bottom panel of figure~\ref{fig:likeMCMC}.  In effect this probability is independent of $f_{\rm mco}$ except at very small values of $f_{\rm mco}$.  This curve marks a formal upper mass boundary of where our $f_{\rm mco}$ limit is valid.

 If $M$ is outside these upper and lower bounds some of the SN light curves should show significant deviations from the standard shape.  Unfortunately there are not good limits on time
dependent deviations in the data because of irregular sampling and intrinsic variations in the light curve shapes. 

It has been shown that, at 95\% confidence, no more than 88\% of the dark matter in the universe can be in compact massive objects with masses greater than $10^{-2}\msun$, closing a gap in the allowed mass for dark matter particles.  
Future data will undoubtedly improve these constraints.  
Gravitational lensing cannot significantly change the conclusions drawn from the high redshift type Ia supernova surveys, but it will be important to account for it in future high precision measurements of the cosmological parameters.

\end{document}